\documentclass[conference]{IEEEtran}
\IEEEoverridecommandlockouts
\usepackage{cite}
\usepackage{amsmath,graphicx}
\usepackage{amssymb,amsfonts}
\usepackage{graphicx}
\usepackage{textcomp}
\usepackage{xcolor}
\usepackage{subfigure}
\usepackage{textcomp} % to use \textperthousand                                                     
\usepackage{siunitx}  
\usepackage{algorithm}
\usepackage{algpseudocode}
\usepackage{xpatch}
\usepackage{pgfplots}
\usepackage{pgfplotstable}
\usepackage{readarray}
\usepackage{siunitx}
\usepackage{bm}
\usepackage[nolist]{acronym}
\usepackage{bbm}
\usepackage{url}

\newcommand{\eyeM}{\bm{\mathrm{I}}}

\newcommand{\bD}{\bm{D}}

\newcommand{\bH}{\bm{H}}

\newcommand{\bv}{\bm{v}}

\newcommand{\cmplx}[1]{\mathbb{C}^{#1}}
\newcommand{\norm}[1]{\|#1\|}
\newcommand{\magn}[1]{|#1|}

\newcommand{\teS}{{\text{S}}}
\newcommand{\teR}{{\text{R}}}
\newcommand{\teD}{{\text{D}}}
\newcommand{\teSR}{{\text{SR}}}
\newcommand{\teSD}{{\text{SD}}}
\newcommand{\teRS}{{\text{RS}}}
\newcommand{\teDS}{{\text{DS}}}
\newcommand{\teDR}{{\text{DR}}}
\newcommand{\teRD}{{\text{RD}}}

\newcommand{\bi}{\bm{i}}
\newcommand{\bb}{\bm{b}}
\newcommand{\ba}{\bm{a}}
\newcommand{\bTheta}{\bm{\Theta}}

\newcommand{\bZ}{\bm{Z}}
\newcommand{\bA}{\bm{A}}

\newcommand{\bS}{\bm{S}}
\newcommand{\bzero}{\bm{0}}
\newcommand{\tG}{{\text{G}}}
\newcommand{\teL}{{\text{L}}}
\newcommand{\tN}{{\text{N}}}
\newcommand{\inve}{{-1}}
\newcommand{\transp}{{\mathrm{T}}}
\newcommand{\im}{\mathrm{j}}
\newcommand{\bzeros}{\bm{0}}

\definecolor{TUMBeamerYellow}    {rgb} {1.000,0.706,0.000}    % RGB 255,180,000
\definecolor{TUMBeamerOrange}    {rgb} {1.000,0.502,0.000}    % RGB 255,128,000
\definecolor{TUMBeamerRed}       {rgb} {0.898,0.204,0.094}    % RGB 229,052,024
\definecolor{TUMBeamerDarkRed}   {rgb} {0.792,0.129,0.247}    % RGB 202,033,063
\definecolor{TUMBeamerBlue}      {rgb} {0.000,0.600,1.000}    % RGB 000,153,255
\definecolor{TUMBeamerLightBlue} {rgb} {0.255,0.745,1.000}    % RGB 065,190,255
\definecolor{TUMBeamerGreen}     {rgb} {0.569,0.675,0.420}    % RGB 145,172,107
\definecolor{TUMBeamerLightGreen}{rgb} {0.710,0.792,0.510}    % RGB 181,202,130

\begin{acronym}
    \acro{AoD}{angle of departure}
    \acro{AoA}{angle of arrival}
    \acro{ULA}{uniform linear array}
    \acro{CSI}{channel state information}
    \acro{LOS}{line of sight}
    \acro{EVD}{eigenvalue decomposition}
    \acro{BS}{base station}
    \acro{MS}{mobile station}
    \acro{mmWave}{millimeter wave}
    \acro{DPC}{dirty paper coding}
    \acro{RIS}{reconfigurable intelligent surface}
    \acro{AWGN}{additive white gaussian noise}
    \acro{MIMO}{multiple-input multiple-output}
    \acro{SISO}{single-input single-output}
    \acro{UL}{uplink}
    \acro{DL}{downlink}
    \acro{OFDM}{orthogonal frequency-division multiplexing}
    \acro{TDD}{time-division duplex}
    \acro{LS}{least squares}
    \acro{MMSE}{minimum mean square error}
    \acro{SINR}{signal to interference plus noise ratio}
    \acro{OBP}{optimal bilinear precoder}
    \acro{LMMSE}{linear minimum mean square error}
    \acro{MRT}{maximum ratio transmitting}
    \acro{M-OBP}{multi-cell optimal bilinear precoder}
    \acro{S-OBP}{single-cell optimal bilinear precoder}
    \acro{SNR}{signal-to-noise ratio}
    \acro{SDR}{Semidefinite Relaxation}
    \acro{SE}{spectral efficiency}
    \acro{GCEs}{Gram channel eigenvalues}
    \acro{LNA}{low-noise amplifier}
    \acro{Tx}{transmit side}
    \acro{Rx}{receive side}
    \acro{HPA}{high power amplifier}
    \acro{LNA}{low noise amplifier}
\end{acronym}

\def\BibTeX{{\rm B\kern-.05em{\sc i\kern-.025em b}\kern-.08em
    T\kern-.1667em\lower.7ex\hbox{E}\kern-.125emX}}
\begin{document}

\title{Physically Consistent Modelling of Wireless Links with Reconfigurable Intelligent Surfaces\\ Using Multiport Network Analysis
%\thanks{Identify applicable funding agency here. If none, delete this.}
}

\author{\IEEEauthorblockN{Josef A. Nossek, \textit{Life Fellow, IEEE}, Dominik Semmler, Michael Joham, and Wolfgang Utschick, \textit{Fellow, IEEE}}
\IEEEauthorblockA{\textit{School of Computation, Information and Technology, Technical University of Munich, 80333 Munich, Germany} \\
email: \{josef.a.nossek,dominik.semmler,joham,utschick\}@tum.de}}

\maketitle

\begin{abstract}
\Acp{RIS} are an emerging technology for engineering the channels of future wireless communication systems.
The vast majority of research publications on \ac{RIS} are focussing on system-level optimization and are based on very simplistic models
ignoring basic physical laws. There are only a few publications with a focus on physical modeling. Nevertheless, the widely employed model is still inconsistent with basic physical laws. 
We will show that even with a very simple abstract model based on isotropic radiators,
ignoring any mismatch, mutual coupling, and losses, each \ac{RIS} element cannot be modeled to simply reflect the incident signal by manipulating its phase only and letting the amplitude unchanged.
We will demonstrate the inconsistencies with the aid of very simple toy examples, even with only one or two \ac{RIS} elements.
Based on impedance parameters, the problems associated with scattering parameters can be identified enabling a correct interpretation of the derived solutions.
\end{abstract}


\begin{figure}[b]
    \onecolumn
    \centering
    \scriptsize{This work has been submitted to the IEEE for possible publication. Copyright may be transferred without notice, after which this version may no longer be accessible.}
    \vspace{-1.3cm}
    \twocolumn
\end{figure}

\section{Introduction}

\subsection{Background and related work}
\acp{RIS} are intended to engineer the propagation environment to improve the performance of wireless communications,
especially in situations where the direct link between the \ac{Tx} and \ac{Rx} is more or less blocked. 
Since many publications about \ac{RIS}  elaborate on why this is an interesting and promising technology \cite{SmartRadioEnvironment}, \cite{SmartWirelessCommunications}, we will not repeat this reasoning here but focus on the physically consistent modeling process and its consequences for the system optimization.

While most research has been performed on system-level optimization, e.g., \cite{SystemSimulator}, 
where over-simplistic models have been used, only a few publications are incorporating basic physical laws in the modeling process \cite{MutualCouplingAware}, \cite{ScatteringRIS}.
Surprisingly, in these two publications, different approaches to modeling have been adopted.
In \cite{MutualCouplingAware}, the analysis is based on impedance parameters, while in \cite{ScatteringRIS}, a scattering parametrization has been chosen.
One can argue which approach is, so to speak, the "better" one.
We will show that this is a matter of taste and may be convenience.
In the impedance parameter approach, the variables are voltages and currents, while in the scattering parameter approach, the variables are incident and reflected waves.
Since these two pairs of variables are simply related to each other by a linear transformation,
the impedance matrices and the corresponding scattering matrices can easily be converted from one to the other.
But interestingly enough, the results given in \cite{MutualCouplingAware} and \cite{ScatteringRIS} finally lead to different conclusions.

Even in the simplest end-to-end \ac{SISO} link with a blocked direct channel between \ac{Tx} and \ac{Rx}, the results in \cite{ScatteringRIS} suggest that the reactive terminations of the \ac{RIS} elements change the phase of the signal received by each element and transmit it to the \ac{Rx} without any change in amplitude, see (41)-(44) in \cite[Section \uppercase\expandafter{\romannumeral 4\relax}]{ScatteringRIS}.
On the other hand, we learn from (12) in \cite{MutualCouplingAware} that any change of that same reactive termination will change the phase and amplitude of the signals sent to the final destination \ac{Rx} simultaneously.
Apparently, it is impossible that both results are true at the same time.
Therefore, we have to go for a careful investigation to find out what is the inherent problem there.

We give one last comment concerning the choice of either scattering or impedance parameters.
For measurements in the microwave range, of course, scattering parameters are to be preferred.
All the available measurement equipment with sound reasons is based on this approach.
But for theoretical derivations, we feel that impedance parameters are better suited and are providing deeper insight, as we will see in the subsequent investigations.

\begin{figure}
\includegraphics{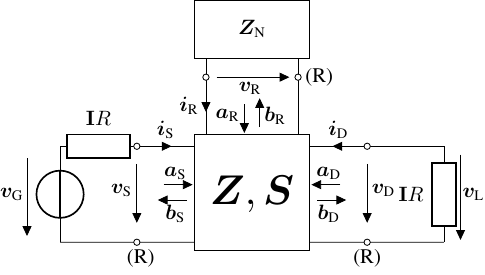}
\caption{Multiport Model}
\label{fig:Three_Port}
\end{figure}

\subsection{Contributions}
We derive the multiport matrix of a \ac{RIS}-aided \ac{MIMO} link based on both
impedance parameters and scattering parameters, neglecting all extrinsic and intrinsic noise sources.
The conversion from impedance to scattering description and vice versa provides insight, especially on the relation of the direct link representation of the two results.
Obviously, both approaches, correctly interpreted, lead to the same final result.
However, there are interesting consequences for the channel estimation process if one chooses to follow the scattering parameter or the impedance parameter approach.

\subsection{Notation}
Bold upper- and lower-case letters denote matrices and column vectors, respectively.
$\bA^\transp$ and $\bA^{\inve}$ denote the transpose and the inverse of matrix $\bA$, respectively.
$\im = \sqrt{-1}$ is the imaginary unit.
Scalars are non-bold letters,
$\magn{a}$, $\arg(a)$ are the magnitude and the phase of a complex scalar $a$.
$\eyeM_n$ is the $n \times n$ identity matrix and $\bzeros$ the all-zeros matrix.

\section{Multiport System Model}
In this section, we firstly describe the basic assumptions about the antenna arrays used at \ac{Tx}, \ac{Rx}, and the \ac{RIS},
that are incorporated in the multiport shown in Figure \ref{fig:Three_Port}. 
This multiport can be characterized by its impedance matrix $\bZ = \bZ^\transp$ (symmetric due to reciprocity), which is partioned as
\begin{equation}
    \begin{bmatrix}
        \bv_\teS\\
        \bv_\teR\\
        \bv_\teD
    \end{bmatrix}
    =
    \begin{bmatrix}
        \bZ_\teS & \bZ_\teSR & \bZ_\teSD\\
        \bZ_\teRS & \bZ_\teR & \bZ_\teRD\\
        \bZ_\teDS & \bZ_\teDR & \bZ_\teD\\
    \end{bmatrix}
    \begin{bmatrix}
        \bi_\teS\\
        \bi_\teR\\
        \bi_\teD
    \end{bmatrix},
\end{equation}
where $\bv_\teS \in 1\si{V} \cdot \cmplx{M} , \bv_\teR \in 1\si{V} \cdot \cmplx{N},  \text{ and } \bv_\teD \in 1\si{V} \cdot \cmplx{K},$ are the complex port voltages at the $M$ \ac{Tx} antennas,
$N$ \ac{RIS} elements and $K$ \ac{Rx} antennas, respectively, and $\bi_\teS, \bi_\teR, \text{ and } \bi_\teD$ are the corresponding port currents.
Equivalently, the multiport can be characterized by 
\begin{equation}
    \begin{bmatrix}
        \bb_\teS\\
        \bb_\teR\\
        \bb_\teD
    \end{bmatrix}
    =
    \begin{bmatrix}
        \bS_\teS & \bS_\teSR & \bS_\teSD\\
        \bS_\teRS & \bS_\teR & \bS_\teRD\\
        \bS_\teDS & \bS_\teDR & \bS_\teD\\
    \end{bmatrix}
    \begin{bmatrix}
        \ba_\teS\\
        \ba_\teR\\
        \ba_\teD
    \end{bmatrix}
\end{equation}
where $\ba_\teS$, $\ba_\teR$, and $\ba_\teD$ are the incident voltage waves and $\bb_\teS$, $\bb_\teR$, and $\bb_\teD$ their reflected counterparts, which are related to the port voltages and currents by%in the following way
\begin{equation}
    \label{eq:VoltageParameterBasic}
    \begin{aligned}
        \bv_{\Gamma} &= \ba_{\Gamma} + \bb_{\Gamma}, \quad &\bi_{\Gamma} &= (\ba_{\Gamma}-\bb_{\Gamma})/R,\\
    \end{aligned}
\end{equation}
where $\Gamma \in \{ \teS,\teR,\teD\}$ refers to each of the ports, and $R$ is the port resistance, assigned to all ports.
Additionally, the wave parameters are given by 
\begin{equation}
    \label{eq:WaveParameterBasic}
    \ba_{\Gamma} =  \frac{1}{2}(\bv_{\Gamma} + R\bi_{\Gamma}), \quad \bb_{\Gamma} =  \frac{1}{2}(\bv_{\Gamma} - R\bi_{\Gamma}).
\end{equation}
We have assumed that the source impedances of the \ac{HPA} outputs, which are connected to the \ac{Tx} antennas,
and the input impedances of the \ac{LNA} inputs,
which are connected as a load to the \ac{Rx} antennas are, without loss of generality, all of the same value $R$.
For the $N$-port characterized by $\bZ_\tN$ (see Fig \ref{fig:Three_Port}), it is assumed that each \ac{RIS} element is terminated with one reactive, i.e., lossless, one-port.
The matrix $\bZ_\tN$, therefore, is diagonal with imaginary entries.

Now we assume that all antenna elements at the \ac{Tx}, \ac{RIS} and \ac{Rx} are isotropic radiators,
although we know that such elements do not exist for vector fields.
However, a field-theoretic justification for using such hypothetical elements is given in \cite{IsotropicRadiators}.
We'd like to note that almost in every system-level publication dealing with antenna arrays, such a model is used.
The self-impedance is real-valued and is assumed to be equal to the port resistance $R$.
Hence, we have power matching between the \acp{HPA} and \ac{Tx}-antennas as well as the \acp{LNA} and \ac{Rx}-antennas.

If no current is flowing across the terminals of an isotropic radiator,
then this radiator does not produce any electromagnetic field, i.e., it is of the so-called canonic minimum scattering type \cite{MinimumScattering}.

The so-called effective area of an isotropic radiator is $A = \frac{\lambda^2}{4 \pi}$ and the gain,
which is uniform for all azimuth and elevation angles is $G=1$.
The ratio between $A$ and $G$ is $\frac{\lambda^2}{4 \pi}$, which is true for any antenna type \cite[Chapter 13.7, pp. 569-575]{ElectromagneticsBook}.

There is, of course, mutual coupling between isotropic radiators, which we can compute as 
\begin{equation}
    z_{21} = \frac{v_2}{i_1} \bigg|_{i_2=0\si{A}} = - \frac{R}{\im kd} \mathrm{e}^{-\im k d},
\end{equation}
where $k=\frac{2\pi}{\lambda}$ is the wave number, $d$ the distance between the antenna excited by current $i_1$
and the second open-circuited antenna,
whose voltage $v_2$ is induced.
Such a mutual impedance is necessary to have communication between \ac{Tx}, \ac{RIS}, and \ac{Rx}.
It exists, of course, also between antenna elements within the \ac{Tx}-array, \ac{RIS} and \ac{Rx}-array
and it will be incorporated by the off-diagonal entries of the matrices $\bZ_\teS$, $\bZ_\teR$, and $\bZ_\teD$.

Now we follow Subsection \uppercase\expandafter{\romannumeral 3\relax}.\textit{B} in \cite{ScatteringRIS} and analyze the \ac{RIS}-aided communication model in Figure \ref{fig:Three_Port} with perfect matching and without intra-array coupling,
resulting in the impedance matrices
\begin{equation}
   \begin{aligned}
        \bZ_\teS &= \eyeM_M R, \quad &\bZ_\teR &= \eyeM_N R, \quad &\bZ_\teD &= \eyeM_K R\\
   \end{aligned}
\end{equation}
and the scattering matrices
\begin{equation}
    \bS_\teS = \bzeros, \quad \bS_\teR=\bzeros, \quad \bS_\teD= \bzeros.
\end{equation}
Because antennas and physical wireless channels are reciprocal,
it is generally true that 
\begin{equation}
    \bZ_\teSR = \bZ_\teRS^\transp, \quad \bZ_\teSD = \bZ_\teDS^\transp, \quad \bZ_\teRD = \bZ_\teDR^\transp.
\end{equation}
However, it is also true that the signal attenuation between \ac{Tx} and \ac{RIS}, between \ac{RIS} and \ac{Rx}, as well as between \ac{Tx} and \ac{Rx} is usually very large.
Hence,
\begin{equation}
    \norm{\bZ_{\teSR}}_{\text{F}} \ll \norm{\bZ_{\teS}}_{\text{F}}, \quad \norm{\bZ_{\teSD}}_{\text{F}} \ll \norm{\bZ_{\teS}}_{\text{F}}, \quad \norm{\bZ_{\teRD}}_{\text{F}} \ll \norm{\bZ_{\teR}}_{\text{F}}
\end{equation}
holds in practice.
Therefore, we keep $\bZ_\teRS$, $\bZ_\teDS$ and $\bZ_\teDR$ as they are and set 
\begin{equation}
    \bZ_\teSR = \bzeros \; \si{\ohm}, \quad  \bZ_\teSD = \bzeros \; \si{\ohm}, \quad  \bZ_\teRD = \bzeros \; \si{\ohm}, 
\end{equation}
which we call the unilateral approximation \cite{TowardCircuitTheory}, \cite{MulitportTheory}.
For the scattering parameters, we have, accordingly, 
\begin{equation}
    \bS_\teSR = \bzeros, \quad  \bS_\teSD = \bzeros , \quad  \bS_\teRD = \bzeros.
\end{equation}
This leads to the multiport impedance matrix
\begin{equation}\label{eq:JointImpedanceModelSimpl}
\bZ = 
    \begin{bmatrix}
        \eyeM R & \bzero & \bzero\\
        \bZ_\teRS & \eyeM R & \bzero\\
        \bZ_\teDS & \bZ_\teDR & \eyeM R\\
    \end{bmatrix}
\end{equation}
as well as the multiport scattering matrix 
\begin{equation}
    \bS = 
    \begin{bmatrix}
        \bzero & \bzero & \bzero\\
        \bS_\teRS & \bzero & \bzero\\
        \bS_\teDS & \bS_\teDR & \bzero\\
    \end{bmatrix}.
\end{equation}
From these assumptions and Figure \ref{fig:Three_Port} we obtain the voltages
\begin{equation}
    \begin{aligned}
        &\bv_\tG = \bv_\teS + \eyeM_M R\bi_\teS, \quad \bv_\teR = -\bZ_\tN \bi_\teR, \quad \bv_\teD = -\eyeM_K R \bi_\teD = \bv_\teL\\
    \end{aligned}
\end{equation}
as well as the scattering parameters
\begin{equation}
    \begin{aligned}
        &\ba_\teS =\frac{1}{2} \bv_\tG, \quad \ba_\teR = \bTheta \bb_\teR,\quad \ba_\teD = \bzeros \si{V}, \quad \bb_\teS = \bzeros\si{V}.
    \end{aligned}
\end{equation}
Additionally, we have
\begin{equation}
    \label{eq:MultiportSimplifiedEqs}
    \begin{aligned}
         \bb_\teD = \bv_\teD = \bv_\teL \quad \text{and} \quad \bTheta = (\bZ_\tN - \eyeM R)(\bZ_\tN + \eyeM R)^{\inve}
    \end{aligned}
\end{equation}
where $\bTheta$ is a diagonal matrix with unit-modulus constraints as $\bZ_\tN$ is assumed to be diagonal and purely imaginary.

From here and since $\bv_\tG = 2\bv_\teS=2R\bi_\teS$ we compute the transfer matrix based on the impedance parameters as
\begin{equation}
    \label{eq:ImpedanceSimplified}
    \bv_\teL = \bD \bv_\tG, \quad \bD = \frac{1}{4R}\left(\bZ_\teDS - \bZ_\teDR(\bZ_\tN + \eyeM R)^\inve \bZ_\teRS \right)
\end{equation}
and based on the scattering parameters as 
\begin{equation}
    \label{eq:ScatteringSimplified}
    \bb_\teD = \bH 2 \ba_\teS, \quad \bH = \frac{1}{2} (\bS_\teDS + \bS_\teDR \bTheta \bS_\teRS).
\end{equation}
From \eqref{eq:MultiportSimplifiedEqs}, we have $\bb_\teD = \bv_\teL$ and $2\ba_\teS = \bv_\tG$. Therefore $\bH$ and $\bD$ must be identical!
We will validate that by using the standard relationships
\begin{equation}
    \bS = (\bZ-\eyeM R) (\bZ + \eyeM R)^\inve \text{ and } \bZ = R(\eyeM + \bS)(\eyeM - \bS)^\inve
\end{equation}
which follow directly from \eqref{eq:VoltageParameterBasic} and \eqref{eq:WaveParameterBasic}.
In particular, with 
\begin{equation}
    (\bZ  - \eyeM R) = \bS (\bZ +\eyeM R),
\end{equation}
it can be shown that
\begin{equation}
    \label{eq:ScatterImpTransitionReflc}
    \begin{aligned}
        &\bS_\teRS = \frac{1}{2R} \bZ_\teRS, \; \bS_\teDR= \frac{1}{2R} \bZ_\teDR \quad \text{and}\\
    \end{aligned}
\end{equation}
\begin{equation}
    \label{eq:ScatterImpTransition}
    \begin{aligned}
        \bS_\teDS = \frac{1}{2R}\left(\bZ_\teDS - \frac{1}{2R}\bZ_\teDR\bZ_\teRS\right).
    \end{aligned}
\end{equation}
It is important to note, that the result in \eqref{eq:ScatterImpTransition} is different from (37) in \cite{ScatteringRIS}.
Using \eqref{eq:ScatterImpTransition} to compare \eqref{eq:ImpedanceSimplified} and \eqref{eq:ScatteringSimplified}, we confirm $\bD = \bH$.
Nevertheless, let us consider the case of a weak or even fully blocked direct link between \ac{Tx} and \ac{Rx}.
From \eqref{eq:ImpedanceSimplified} we get with $\bZ_\teDS = \bzeros \si{\ohm}$,
\begin{equation}
    \label{eq:ImpedanceSimplifiedBlocked}
    \bD_0 = \bD|_{\bZ_\teDS = \bzeros \si{\ohm}} = - \frac{1}{4R} \bZ_{\teDR}(\bZ_\tN + \eyeM R)^\inve \bZ_\teRS,
\end{equation}
while setting $\bS_\teDS = \bzeros$ in \eqref{eq:ScatteringSimplified} would lead to 
\begin{equation}
    \label{eq:ScatteringSimplifiedWOConst}
    \bH|_{\bS_\teDS = \bzeros} = \frac{1}{2} \bS_\teDR \bTheta \bS_\teRS,
\end{equation}
which obviously is different from $\bD_0$!
However, by using \eqref{eq:ScatterImpTransition}, we get
\begin{equation}
    \label{eq:ScatterConstOffset}
    \bS_\teDS^0= \bS_\teDS |_{\bZ_\teDS = \bzeros  \si{\ohm}} = -\bS_\teDR \bS_\teRS.
\end{equation}
Substituting \eqref{eq:ScatterConstOffset} into \eqref{eq:ScatteringSimplified}, we obtain
\begin{equation}
    \begin{aligned}
    \bH_0 &= \frac{1}{2} (- \bS_\teDR \bS_\teRS + \bS_\teDR \bTheta \bS_\teRS) = \frac{1}{2} \bS_\teDR(\bTheta - \eyeM_N)\bS_\teRS\\
    &= \frac{1}{2} \bS_\teDR \Big((\bZ_\tN - \eyeM R)(\bZ_\tN + \eyeM R)^{\inve} \\
    &\quad-(\bZ_\tN + \eyeM R)(\bZ_\tN + \eyeM R)^{\inve}\Big)\bS_\teRS\\
    &=-\frac{1}{2} \bS_\teDR \left(2 \eyeM_N R(\bZ_\tN + \eyeM R)^\inve\right) \bS_\teRS\\
    &= -\bS_\teDR R(\bZ_\tN + \eyeM R)^\inve \bS_\teRS\\
    &= -\frac{1}{4R} \bZ_\teDR (\bZ_\tN + \eyeM R)^\inve \bZ_\teRS = \bD_0,
\end{aligned}
\end{equation}
which is now correct and identical to $\bD_0$ from \eqref{eq:ImpedanceSimplifiedBlocked}.

Therefore, the matrix in between $\bS_\teDR$ and $\bS_\teRS$ is not only changing the phase of the signals emitted from each \ac{RIS} element, 
but simultaneously, the amplitude is changed.
This important consequence has been overlooked in \cite{ScatteringRIS} and many other publications are based on the assumption in (37) of \cite{ScatteringRIS},
which erroneously states that $\bS_\teDS = \frac{1}{2R} \bZ_\teDS$. A careful conversion between $\bS$ and $\bZ$ matrices shows the correct result in \eqref{eq:ScatterImpTransition}.

There is still the question why the intuitive reasoning in \cite{ScatteringRIS} is wrong.
This is because in \eqref{eq:ScatteringSimplifiedWOConst} the salient assumption is that only the waves $\ba_\teR$,
which are reflected by the reactive terminations,
are propagating toward the final destination, i.e. the \ac{Rx}.
But at the ports of the \ac{RIS} elements,
we have not only $\ba_\teR$, but the superposition of $\ba_\teR$ and $\bb_\teR$ simultaneously and therefore,
this superposition of both is propagated towards the \ac{Rx}. 
But the contribution of $\bb_\teR$, which is not incorporated in the triple product $\bS_\teDR \bTheta \bS_\teDS$
has been moved to the element $\bS_\teDS$ and has been overlooked there.

For the problem of channel estimation, we have to recognize that the three channel matrices $\bZ_\teDS$, $\bZ_\teRS$, and $\bZ_\teDR$ are truly different and independent entities.
The statistical fading processes of their random entries are uncorrelated and even statistically independent.
On the contrary, if one tries to estimate $\bS_\teDS, \bS_\teDR$ and $\bS_\teRS$, it is important to note the deterministic dependency of $\bS_\teDS$ on $\bS_\teRS$ and $\bS_\teDR$.

If somebody wants to use the entries of $\bTheta$ as optimization parameters instead of the entries of $\bZ_\tN$,
we can use the impedance parameter description equally well
\begin{equation}
    \bD = \frac{1}{4R}\left(\bZ_\teDS - \frac{1}{2R}\bZ_\teDR \bZ_\teRS + \frac{1}{2R} \bZ_\teDR \bTheta \bZ_\teRS\right).
\end{equation}

Summarizing the two main messages obtained from the above derivation, we have:
\begin{itemize}
    \item A change of the lossless terminations of the \ac{RIS} elements do not only change the phase of the signals propagating from the \ac{RIS} to the \ac{Rx}.
The amplitude of said signals will unavoidably also be changed.
    \item Even when the direct link betweeen \ac{Tx} and \ac{Rx} is completely blocked, the scattering matrix $\bS_\teDS = -\bS_\teDR \bS_\teRS$ is not equal to the zero matrix.
\end{itemize}
Next, we will show with extremely simple scenarios that ignoring the effect of the lossless terminations on the signal amplitudes is quite substantial.

\section{Numerical Results}

\subsection{Single RIS Element}
First, we will demonstrate the amplitude change with a \ac{LOS}-\ac{SISO} link with a single \ac{RIS} element.
We have $M=N=K=1$, and we assume a blocked direct link.
In that scenario, we can easily observe that the amplitude of the signal received by the \ac{Rx} will be quite different for different lossless terminations.
We have 
\begin{equation}
    \begin{aligned}
       &    z_\teS=R, \; z_\teD=R,\; z_\tN=\im X, z_\teDS=0\; \si{\ohm},\\
       & z_\teRS = \im \frac{R \lambda}{2 \pi d_\teRS} e^{-\im 2 \pi \frac{d_\teRS}{\lambda}}, \; z_\teDR = \im \frac{R \lambda}{2 \pi d_\teDR} e^{-\im 2 \pi \frac{d_\teDR}{\lambda}}
    \end{aligned}
\end{equation}
and we choose $d_\teRS = 10^2 \lambda$ and $d_\teDR = 10^3 \lambda$.
This leads to
\begin{equation}
    D_0 = \frac{1}{(4\pi)^2} 10^{-5} \frac{R}{R+\im X}.
\end{equation}
We normalize $D_0$ to take out the path loss and get 
\begin{equation}
    D_0^\prime =  (4\pi)^2 10^{5} D_0 = \frac{1}{1+\im x}, \quad x = \frac{X}{R}
\end{equation}
as well as
\begin{equation}
    \phi = \arg D_0^\prime = -\arctan x \; \text{ and } \; \magn{D_0^\prime} = (1+x^2)^{-\frac{1}{2}}.
\end{equation}
We summarize the results for a number of different terminations in Table \ref{tab:SingleEl}.
\begin{table}[h!]
    \centering
    \begin{tabular}{ |c|c|c|c| } 
     \hline
     $x$ & $\phi/\circ$ & $\magn{D_0^\prime}$&$10\log_{10}\magn{D^\prime_0}^2/$\si{dB} \\ 
     \hline
     $-\infty$ & $90$ & $0$ & $-\infty$\\ 
     $-1$ & $45$ & $\frac{1}{\sqrt{2}}$ &$-3$\\ 
     $0$ & $0$ & $1$ &$0$\\ 
     $1$ & $-45$ & $\frac{1}{\sqrt{2}}$ &$-3$\\ 
     $\infty$ & $-90$ & $0$ &$-\infty$\\ 
     \hline
    \end{tabular}
    \vspace{0.1cm}
    \caption{Single RIS Element}
    \label{tab:SingleEl}
    \end{table}
There is quite a dramatic change in amplitude associated with the phase shift.
Although in this example, any other phase shift than $\phi = 0^{\circ}$ does not make any sense in the case of a single \ac{RIS} element, this example already demonstrates the importance of the correct interpretation of the model.

\subsection{Two RIS Elements}
\begin{figure}[h!]

	\centering
	\includegraphics[scale=0.8]{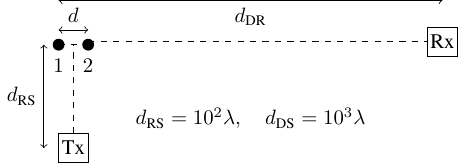}
	\caption{\ac{SISO} Link with Two-Element \ac{RIS}}
	\label{fig:Scenario}
	
\end{figure}

Now we add a second element according to Figure \ref{fig:Scenario}.
We normalize $D_0$ as before and get 
\begin{equation}
    D_0^\prime = \frac{1 + \im x_2 + e^{-\im 2\pi \frac{d}{\lambda}} (1+\im x_1) }{(1+\im x_1)(1+\im x_2)},
\end{equation}
with $x_1 = \frac{X_1}{R}$ and $x_2 = \frac{X_2}{R}$ being the normalized reactances terminating the two \ac{RIS} elements.
Now we maximize $\magn{D^\prime_0}$ to have the best possible signal at the \ac{Rx}.
In Table \ref{tab:TwoEl} we display the optimum reactances and the corresponding values of the normalized magnitude of the transfer function.
\begin{table}[h!]
    \centering
    \begin{tabular}{ |c|c|c|c|c| } 
     \hline
     $d$&$x_1$&$x_2$ &  $\magn{D_0^\prime}^2$&$10\log_{10}\magn{D^\prime_0}^2/$\si{dB} \\ 
     \hline
        $0$ & $0$ & $0$ & $4$ & $6$\\
        $\frac{\lambda}{4}$ & $\sqrt{2}-1$ & $1-\sqrt{2}$ & $\frac{1}{6-4\sqrt{2}}$ & $4,6$ \\
        $\frac{\lambda}{2}$ & $1$ & $-1$ & $1$ & $0$\\
        $\frac{3\lambda}{4}$ & $1-\sqrt{2}$ & $\sqrt{2}-1$ & $\frac{1}{6-4\sqrt{2}}$ & $4,6$ \\
        $\lambda$ & $0$ & $0$ & $4$ & $6$\\
     \hline
    \end{tabular}
    \vspace{0.1cm}
    \caption{Optimum Design for the \ac{SISO} link with a two-element RIS}
    \label{tab:TwoEl}
    \end{table}
It is important to note that an optimization based on section  \uppercase\expandafter{\romannumeral 4\relax}.\textit{A} of \cite{ScatteringRIS} leads to a completely different result.
The receive signal power would have been independent of the distance $d$ of the two \ac{RIS} elements constant at $0$ \si{dB}.

It is important to note that the optimization result for this conventional model has a degree of freedom, i.e. one of the elements of the diagonal $\bTheta$ can be set to any arbitrary value, e.g. zero, without changing the achieved power gain.
But this is true only, if the diagonal $\bTheta$ is applied to a system, which faithfully follows the conventional, but erroneous model \eqref{eq:ScatteringSimplifiedWOConst}.
If this $\bTheta$ is appled to to a more realistic model, i.e. the physically consistent one \eqref{eq:ImpedanceSimplifiedBlocked}, then the previously arbitrary choice is not arbitrary anymore and has quite an influence on the achieved power gain.

\begin{figure}%[!ht]%[!t]

	\centering

    \includegraphics{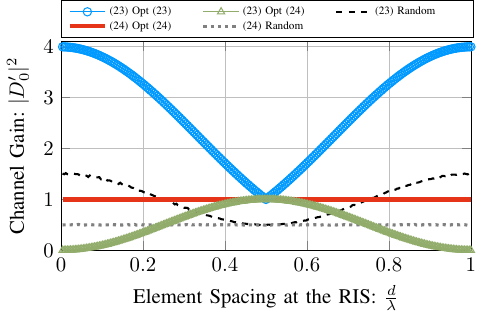}

	\caption{Model Comparison over Element Spacing}
\label{fig:ElementSpacing}

\end{figure}

The whole picture is shown in Figure \ref{fig:ElementSpacing}, where the curve with the blue circles show the achievable power gain as a function of the spacing between the two RIS elements. Here the optimization has been carried out on the physically consistent model \eqref{eq:ImpedanceSimplifiedBlocked} showing quite an improved performance compared to the conventional model, which is given by the red curve. In contrast, applying the optimization result obtained with the conventional model to physcally consistent one, we get an even worse power gain given by the curve with the green triangles. Here we have set the phase factor for the RIS element 1 to zero degrees. With other choices we can generate very many different curves.

As a baseline comparison there have been added two more curves, which are not optimization results but have been generated by averaging over many independently and uniformly distributed random phase realizations. Obviously, these results, given by dashed and dotted lines, are well below their optimized counterparts.
It is interesting to note that now the gain $\magn{D_0^\prime}^2$ depends on the choice of the phase applied for the phase $\phi_1$ of \ac{RIS} element numbered one.
In the conventional model \eqref{eq:ScatteringSimplifiedWOConst}, this choice would have no effect on the gain, but in reality, it does.

\section{Conclusion}
The discrepancy of two different models for a \ac{RIS}-aided communication link given in \cite{MutualCouplingAware} and \cite{ScatteringRIS} has been resolved.
This difference is most striking in the case of a blocked direct link,
but even if this is not the case,
we have to be careful in interpreting the different approaches in describing the same scenario correctly.

The important consequence of correct modeling has been demonstrated with a very simple toy scenario and 
by assuming isotropic radiators by ignoring intra-array mutual coupling.

What has been called physically consistent modelling was simply the correct separation of the direct link from the link via the \ac{RIS}. There are a number of further steps ahead to achieve physial consistency in a broader sense.
Obviously, future research has to focus on real-world scenarios with a large number of \ac{RIS} elements,
including intra-array mutual coupling, mismatch, and losses.
For taking into account finite bandwidth, we must also proceed with real-world antenna elements instead of isotropic ones.

The fact that there are three different channel matrices,
the one from \ac{Tx} to the \ac{RIS}, the one from the \ac{RIS} to the \ac{Rx}
and eventually a direct link from \ac{Tx} to \ac{Rx},
have to be estimated, should be reconsidered.
Whether this is based on impedance parameters or scattering parameters can make a difference.
The same is true for expected values of achievable rates taking into account the statistical properties of the different channel matrices.
\bibliographystyle{IEEEtran}
\bibliography{refs}
\end{document}